\documentclass[prb,twocolumn,,floatfix]{revtex4}
\usepackage{graphicx}
\usepackage{bm}
\usepackage{amssymb}
\usepackage{color}

\newcommand{\be}{\begin{equation}}
\newcommand{\ee}{\end{equation}}

\def \ua{{\uparrow}}
\def \da{{\downarrow}}
\def \be{\begin{equation}}
\def \ee{\end{equation}}
\def \ba{\begin{array}}
\def \ea{\end{array}}
\def \bea{\begin{eqnarray}}
\def \eea{\end{eqnarray}}
\def \nn{\nonumber}
\def \half{{1\over 2}}
\def \etal{{\it {et al}}}

\def \br{{\bf r}}

\def \bv{{\bf v}}
\def \bq{{\bf q}}
\def \bk{{\bf k}}

\def \a{{\alpha}}
\def \t{{\theta}}

\def \D{{\Delta}}
\def \d{{\delta}}
\def \w{{\omega}}
\def \s{{\sigma}}

\def \h{{\eta}}

\def \z{{\zeta}}

\def \nd{{^{\vphantom{\dagger}}}}
\def \yd{^\dagger}
\def \av#1{{\langle#1\rangle}}

\begin{document}

\title{Evolution of the Fermi surface in phase fluctuating d-wave
superconductors}
\author{Erez Berg$^{1,2}$ and Ehud Altman$^1$}
\affiliation{$^1$ Department of Condensed Matter Physics, Weizmann Institute of Science,
Rehovot, 76100, Israel\\
$^2$ Department of Physics, Stanford University, Stanford CA 94305-4045, USA }

\begin{abstract}
One of the most puzzling aspects of the high $T_c$ superconductors
is the appearance of Fermi arcs in the normal state of the
underdoped cuprate materials. These are loci of low energy
excitations covering part of the fermi surface, that suddenly
appear above $T_c$ instead of the nodal quasiparticles. Based on a
semiclassical theory, we argue that partial Fermi surfaces arise
naturally in a d-wave superconductor that is destroyed by thermal
phase fluctuations. Specifically, we show that the electron
spectral function develops a square root singularity at low
frequencies for wave-vectors positioned on the bare Fermi surface.
We predict a temperature dependence of the arc length that can
partially account for results of recent angle resolved photo
emission (ARPES) experiments.

\end{abstract}

\date{\today}
\maketitle

The transition from a superconductor to a normal state at $T_c$
can take place in two qualitatively different ways. Conventional
metallic superconductors take the BCS-Eliashberg route by which
the amplitude of the complex order parameter
$|{\Delta}|e^{i{\varphi}}$ vanishes at $T_c$ together with the
quasi-particle gap. Thus the transition is accompanied by
formation of a normal Fermi surface. Phase fluctuations play a
negligible role
in this transition and $T_c$ is determined by the zero temperature gap ${%
\Delta}_0$. On the other hand, in superconductors with small
superfluid density $n_s$, the phase-stiffness defines a
temperature scale $T_\t\sim n_s/m^\star$ that may be much lower
than ${\Delta}$. The transition then occurs at $T_c\sim T_\t$ due
to disordering of the phase by thermal fluctuations. This is
thought to be the case in the underdoped cuprate
superconductors\cite{Uemura89,EmeryKivelson95}, where $T_c$ is
seen to increase with doping concomitantly with the superfluid
density\cite{Uemura89} and is anti correlated with the
quasi-particle gap. Is such a transition, driven by phase
fluctuations, accompanied by formation of a Fermi surface as is
the BCS case?

Naively, the answer is negative. At temperatures not too high above $T_c$ ($%
T<T_{MF}\sim {\Delta}_0$) pairing survives in the normal state. One might
expect that in this regime the single particle spectral function remains
peaked at finite frequency of order $\Delta_0$, the energy needed to
dissociate a pair. This reasoning is certainly correct for an s-wave
superconductor with clear separation of scales between the phase stiffness
and the larger quasi-particle gap. Does it still hold in a d-wave
superconductor, where such a clear separation of scales cannot exist because
of the vanishing gap at the nodal points?

In this paper we investigate the evolution of the Fermi surface in the
normal state of a $d$-wave superconductor by studying the coupling between
nodal quasi-particles and strong phase fluctuations. Our main result is that
for any value of the anti-nodal gap a sharp Fermi surface forms in the
normal state immediately above $T_c$. Nevertheless this is not a normal
Fermi liquid. Because of coupling to thermal phase fluctuations, the
quasi-particle peak on the Fermi surface is replaced by a square root
singularity in the spectral function $A(\mathbf{k}=\mathbf{k}_f,{\omega})= a(%
\mathbf{k})/\sqrt{{\omega}}$. This singularity is present everywhere on the
Fermi surface, but at wave-vectors far from the gap nodes, that is ${\delta}
\mathbf{k}>\sqrt{{\langle \nabla{\varphi}^2\rangle}}\equiv q_0(T)$, it is
exponentially suppressed by the pre-factor $a(\mathbf{k})$. Therefore a peak
at zero frequency is observable in practice only along arcs of length $\sim
q_0$ on the Fermi surface, centered at the nodal points. The dominant
spectral feature further from the node, is a broad peak at frequency ${\omega%
}={\Delta}_\bk$.

Broadening of the quasi-particle peak at ${\omega}={\Delta}_\bk$ by thermal
phase fluctuations in the normal state was noted in previous analyses \cite%
{FranzMillis,Dorsey}. However the appearance of a singularity at zero
frequency signaling the formation of a Fermi surface, was overlooked. We
argue that this physics underlies the formation of the Fermi arcs seen in
angle resolved photoemission spectroscopy (ARPES) of underdoped cuprates\cite%
{Norman98,Kanigel06}.

Before proceeding we note that a number of recent works proposed
alternative mechanisms for the formation of Fermi arcs and their
evolution with temperature. Varma and Zhu \cite{Varma} have argued
that current fluctuations in a state with broken time reversal
symmetry can explain the observed phenomena. Paramekanti and Zhao
\cite{ParamekantiZhao} postulated that the normal state of the
underdoped cuprates is dominated by a quantum critical point with
deconfined spinons at the critical doping level for onset of
superconductivity. In their theory, the broadening of the spectral
peaks accompanying the deconfinement leads to formation of the
arcs. Kim \emph{et al}\cite{EunAh} have discussed the formation of
Fermi arcs due to critical fluctuations of an underlying nematic
order parameter in the presence of quenched disorder.

For concreteness and to make contact with experiments in the cuprates, we
consider a layered structure relevant to these materials. The main
ingredient in our analysis, leading to emergence of a Fermi surface, is the
coupling of the dirac quasi-particles to the local currents $\mathbf{j}_s=
n_s\nabla {\varphi}$ induced by the phase fluctuations. These currents are
conveniently split into two contributions with markedly different effect on
the quasiparticles: A transverse part originating from vortex configurations
and an irrotational, or longitudinal part. We find that the interesting
effects on the single particle spectral function result from the transverse
fluctuations. To simplify the subsequent analysis we treat the vortices as
purely thermal fluctuations and neglect their dynamics altogether. This is a
reasonable assumption in the normal state where vortex motion is expected to
be diffusive and slow on electronic time scales.
The longitudinal fluctuations, by contrast, can be shown to be highly
quantum \cite{Randeria}. However, we will show that their coupling to the
quasi-particles is irrelevant, even if the small gap in their dispersion,
due to the c-axis Josephson plasmon, is neglected.

Under these assumptions, the electron Green's function relevant to
ARPES is given by:
\begin{equation}
G^<(\mathbf{r},t)=\int\mathcal{D} \mathbf{j}_\perp(\mathbf{r}) P\left[%
\mathbf{j}_\perp(\mathbf{r})\right] {\langle {\psi^{\dagger}(0,0)}\psi(\mathbf{r}%
,t)\rangle}_{\{\mathbf{j}_\perp (\mathbf{r})\}}.  \label{G}
\end{equation}
The external integral implements a thermal average over the
transverse currents (or vortex configurations). The thermal weight
of a current configuration, $P[\mathbf{j}_\perp(\mathbf{r})]$, is
determined by a classical $x-y$ model\cite{scalapino}. The internal
average is the Green's function in a given static current
configuration. This includes integration over the dynamical
longitudinal phase fluctuations.

\begin{figure}[t]
\centering \includegraphics[scale=0.37]{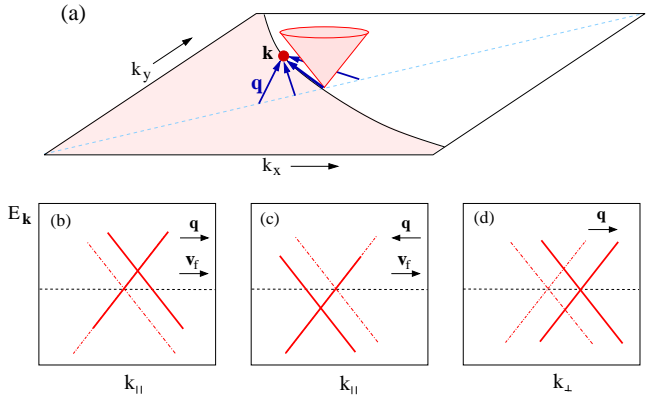}
\caption{In the presence of
static current, the dirac cone, seen in panel (a) on the bare Fermi surface,
is shifted in energy and boosted by the wave-vector
$\bq$ associated with the super-current. If the current is parallel to $\bv_f$
as in (b,c) the net effect is that the cone slides up or down leaving one point
of zero energy excitations invariant. If $\bq\perp \bv_f$ as in (d)
the nodal point is transported along the bare Fermi surface. As shown
in panel (a) there is a continuous range of currents $\bq$ that lead to a zero energy
peak on each point of the bare Fermi surface. This pile up of spectral weight results in a singular
peak at zero energy on the Fermi surface.}
\label{fig:stat}
\end{figure}

In the spirit of refs. [\onlinecite{FranzMillis,Volovik}] we will
approximate (\ref{G}) by an average over a distribution of uniform
currents. This amounts to a semiclassical approximation valid if
spatial variations of the current over a Fermi wavelength are
small (i.e. $k_f\xi \gg 1$). Interference effects due to coherent
scattering of quasi-particles by vortices are neglected in this
scheme, which is justified at the temperatures of interest in the
normal state.

With these approximations, the spectral function on the two dimensional
plane is given by:
\begin{equation}
A(\mathbf{k},{\omega})=\int d^2 q P(\mathbf{q}) A_{\mathbf{q}}(\mathbf{k},{%
\omega}).  \label{Akw}
\end{equation}
Here $A_{\mathbf{q}}(\mathbf{k},{\omega})$ is the spectral function in the
presence of a uniform supercurrent parameterized by a wave-vector $\mathbf{q}
$. Such current corresponds to a pairing field that depends on the center of
mass coordinate of the pair as ${\Delta}_\bk e^{i2\mathbf{q}\cdot\mathbf{R}}$%
.

We may now use the standard BCS formalism in order to calculate $A_{\mathbf{q%
}}(\mathbf{k},{\omega})$ for a given supercurrent $\mathbf{q}$. This is
easily seen to be
\begin{equation}
A_{\mathbf{q}}(\mathbf{k},{\omega})=|u_{\mathbf{k}-\mathbf{q}}|^2{\delta}({%
\omega}-E_{\mathbf{k}-\mathbf{q}}(\mathbf{q})) +|v_{\mathbf{k}-\mathbf{q}}|^2%
{\delta}({\omega}+E_{\mathbf{q}-\mathbf{k}}(\mathbf{q}))  \label{Aq}
\end{equation}
and the quasi-particle dispersion to leading order in $\mathbf{q}$ is given
by
\begin{equation}
E_\bk(\mathbf{q})\approx\mathbf{v}_f(\mathbf{k})\cdot\mathbf{q}+\sqrt{%
\xi_\bk^2+{\Delta}_\bk^2}.  \label{Eq}
\end{equation}
for $\mathbf{k}$ near the bare Fermi surface, defined by $\xi_{\mathbf{k}_f}=%
{\epsilon}_{\mathbf{k}_f}-\mu=0$. To this order in $\mathbf{q}$, the
coherence factors $u_\bk$ and $v_\bk$ are unchanged by the current, that is $%
|u_\bk|^2=(1+\xi_\bk/\sqrt{\xi_\bk^2+{\Delta}_\bk^2})/2$, and $%
|v_\bk|^2=1-|u_\bk|^2$.

The essential physics leading to emergence of a Fermi surface in the normal
state can be understood qualitatively from Eqs. (\ref{Aq}) and (\ref{Eq}).
Fig. \ref{fig:stat} shows the dispersion of the quasi-particle peaks in $%
A_\bq(\mathbf{k},{\omega})$ for different directions of the supercurrent.
The momentum of the peak is shifted by $\mathbf{q}$ because of the boost
associated with the supercurrent, but its frequency is also "Doppler
shifted" by $\mathbf{v}_f\cdot\mathbf{q}$. For a current parallel (or
anti-parralel) to $\mathbf{v}_f$, this amounts to the dirac cone sliding up
(or down) on the slope of the original dispersion as seen in Fig. \ref%
{fig:stat}(b,c). The important thing to note is that in both cases there is
a zero frequency peak, or "Fermi point", whose position is independent of
the current. On the other hand, for current perpendicular to $\mathbf{v}_f$
there is no doppler shift, and the node is simply transported a distance $%
\mathbf{q}$ along the Fermi surface (Fig. \ref{fig:stat}d). In general an
arbitrary current $\mathbf{q}$ will always lead to a zero frequency peak
located precisely on the bare Fermi surface. The parallel component of $%
\mathbf{q}$ leaves the zero frequency peak in place on the nodal point,
while the perpendicular component transports it on the fermi surface. In
fact, as shown in Fig. \ref{fig:stat}a, for every point on the Fermi
surface, there is a continuous range of wave-vectors $\mathbf{q}$ that lead
to a zero frequency peak at that point. This pile up is the origin of the
zero frequency singularity that develops on the Fermi surface. However, the
probability for a current fluctuation with wavevector larger than a typical $%
q_0$ is strongly suppressed in the distribution $P(\mathbf{q})$. Therefore
on points of the fermi surface at a distance larger than $q_0$ from the node
the singularity is multiplied by a very small pre-factor making it
essentially unobservable.

We now turn to a systematic derivation of the spectral function using Eqs. (%
\ref{Akw}),(\ref{Aq}) and (\ref{Eq}). We concentrate on wavevectors $\mathbf{%
k}$ sufficiently close to the nodes so that we may approximate the
dispersion by a Dirac cone. Namely: ${\Delta}_\bk=\mathbf{v}_\D\cdot
(\mathbf{k}-\mathbf{k}_n)$ and $\xi_\bk=\mathbf{v}_f\cdot
(\mathbf{k}-\mathbf{k}_n)$. Let us also change variables
from $\mathbf{q}$ to ${\eta}=\mathbf{v}_f\cdot\mathbf{q}$ and $\zeta=\mathbf{%
v}_\D\cdot\mathbf{q}$. Plugging these definitions to (\ref{Akw}) we have
\begin{widetext}
\bea
A(\bk,\w)={1\over 2}\int d\h d\z P(\h,\z)&&\sum_{s=\pm 1}\left(1+s{\xi-\h\over\sqrt{(\xi-\h)^2
+(\D-\z)^2}}\right)\d(\w-\h- s\sqrt{(\xi-\h)^2+(\D-\z)^2})
\label{Aintegral}
\eea
\end{widetext}
where for notational simplicity we omitted the subscript $\mathbf{k}$ from ${%
\Delta}_\bk$ and $\xi_\bk$. The distribution $P({\eta},{\zeta})$
is obtained by a simple change of variables from $P(\mathbf{q})$,
which is taken to be a gaussian of the form 
$P(\mathbf{q})=\mathcal{N}\exp(-\frac{\mathbf{q}^2}{q_0^2(T)})$[\onlinecite{FranzMillis}].
We will later discuss the temperature dependence of $q_0$.

In Ref. [\onlinecite{FranzMillis}], the integral (\ref{Aintegral}) was
approximated by expanding to leading order in ${\eta}/{\Delta}_\bk$ and ${%
\zeta}/{\Delta}_\bk$. For $\mathbf{k}$ on the Fermi surface far from a gap
node this approximation does give the the dominant spectral feature, which
in this region is simply a broadened peak at frequency ${\omega}\sim {\Delta}%
_\bk$. However, the expansion cannot be generally valid for a superconductor
with gap nodes, on which these ratios diverge. In particular, it cannot
capture the pile up of low energy spectral weight on the Fermi surface due
to movement of the Dirac cone in energy and momentum, as illustrated in Fig. %
\ref{fig:stat}.
\begin{figure}[b]
\centering \includegraphics[scale=0.5]{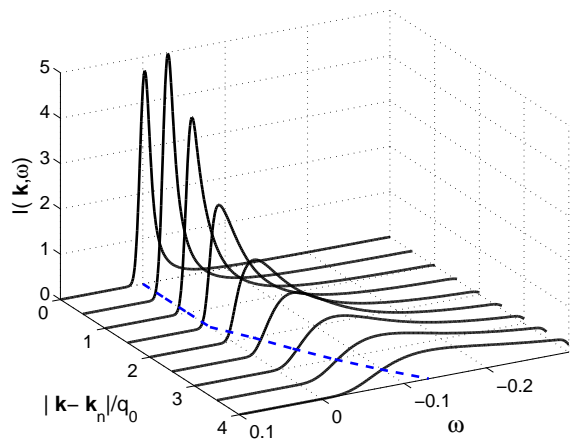}
\caption{Calculated ARPES energy distribution curves (EDC) for different wave-vectors $\bk$ on the Fermi surface
at increasing distance from the node. The curves are obtained by numerical integration
of Eq. (\ref{AintD}), multiplied by the Fermi function and convolved
with a narrow gaussian function to mimic the finite experimental resolution.
The dashed line follows the dispersion of the peak, demonstrating the formation
of a Fermi arc. Evidently the peak departs from
$\w=0$ only at $\bk-\bk_n\approx 1.5 q_0$.}
\label{fig:ARPES_D}
\end{figure}

We proceed to evaluate (\ref{Aintegral}) without resorting to an expansion
in ${\eta}/{\Delta}_\bk$ and ${\zeta}/{\Delta}_\bk$. In what follows we
outline the main steps in the derivation and relegate the details to the
online supplementary information. First we transform to the polar
coordinates $(r,{\theta})$ defined by $r=\sqrt{(\xi-{\eta})^2+({\Delta}-{\zeta%
})^2}$ and $\cos{\theta}=(\xi-{\eta})/r$. This allows to resolve the delta
functions in (\ref{Aintegral}) by integrating over $r$. Next to facilitate
the angular integration we make another change of variables via ${\alpha}%
=s\cos{\theta}$. Now, for
wavevectors $\mathbf{k}$ on the Fermi surface (i.e. $\xi=0$) the integral
reduces to a rather simple form:
\begin{equation}
A(\mathbf{k},{\omega})={\frac{|{\omega}|}{2}}\int_{-1}^1 d{\alpha}{\frac{(1-{%
\alpha})^{1/2}}{(1+{\alpha})^{5/2}}}F(|{\omega}|,{\alpha})  \label{AintD}
\end{equation}
where
\begin{eqnarray}
F(|{\omega}|,{\alpha})={\mathcal{N}} \exp\left[{\frac{-|{\omega}|^2{\alpha}^2}{%
v_f^2 q_0^2(1+{\alpha})^2}} -{\frac{\left(|{\omega}|\sqrt{\frac{1-{\alpha} }{%
1+{\alpha}}}+{\Delta}\right)^2}{v_\D^2 q_0^2}}\right]  \label{F0D}
\end{eqnarray}
and ${\mathcal{N}}\equiv 1/( \pi v_f v_\D q_0^2)$. For $|{\omega}|\ne 0$,
the divergence at ${\alpha}=-1$ in the integrand of (\ref{AintD}) is cut off
by the first term in the exponent in (\ref{F0D}), which effectively
introduces a new lower bound to the integral. The singular contribution to
the spectral function is now easily evaluated as:
\begin{equation}
A_{sing}(\mathbf{k},{\omega})=A_0(v_\D q_0)^{-1}e^{-{\Delta}_\bk^2/v_\D^2 q_0^2}%
\sqrt{\frac{v_f q_0}{|{\omega}|}}  \label{Asing}
\end{equation}
which is the advertised zero frequency singularity. $A_0$ is a
constant of order unity. Note that the singular contribution comes
from ${\alpha}=-1$, which in the original variables
corresponds to the range of currents $\mathbf{q}$ defined by the condition $%
\mathbf{q}\cdot \mathbf{v}_\D={\Delta}_\bk$. This is precisely the condition
for wave-vectors $\mathbf{q}$ illustrated in Fig. \ref{fig:stat}a that pile
up low frequency spectral weight on the point $\mathbf{k}$ on the Fermi
surface. If $\mathbf{k}$ is far from the node compared to a typical current
fluctuation $q_0$, then ${\Delta}/v_\D q_0$ is large and the singularity is
strongly suppressed by the exponential pre-factor in (\ref{Asing}). Then the
dominant spectral feature is a broad peak at frequency ${\omega}\sim{\Delta}$
obtained from taking the saddle point of (\ref{AintD}).

The ARPES spectrum is related to the spectral function via $I(\mathbf{k},{%
\omega})=n_F({\omega})A(\mathbf{k},{\omega})$. In Fig. \ref{fig:ARPES_D} we
plot this function for $\mathbf{k}$ points on the Fermi surface at varying
distance from the node. The curves are obtained by numerical integration of (%
\ref{AintD}) and the convolution with a narrow gaussian function
which mimics the experimental resolution. The overall picture is
appealingly similar to what is seen in ARPES experiments in the
normal state of under doped cuprates\cite{Norman98,Kanigel06}. It
is seen that the sharp peaks are pinned to zero frequency along a
line on the Fermi surface up to a critical distance from the node,
which is of the order of $q_0$. At larger distance from the node
the peak departs from zero frequency, quickly approaches
${\omega}={\Delta}_\bk$ and broadens considerably.

A similar analysis as outlined above can be carried out for points $\mathbf{k%
}$ on the line of nodes ${\Delta}_\bk=0$ away from the Fermi surface. In
this case the singular contribution to the spectral function is given by
\begin{equation}
A_{sing}(\mathbf{k},{\omega})=A_0(v_\D q_0)^{-1}\sqrt{\frac{v_f q_0}{|\xi-{%
\omega}|}},
\end{equation}
which appears as a sharp peak dispersing as $\xi_\bk$. This too is in
qualitative agreement with ARPES.

We turn to discuss the nature of the phase field in the normal state and the
contribution of different types of fluctuations to the electron spectral
function. The phase field of a fluctuating 2d superconductor can be
decomposed into a longitudinal (non singular) part and a transverse, or
vortex contribution. These components are governed by independent actions in
the long wavelength limit.

\emph{Longitudinal phase fluctuations --} These fluctuations may
be treated at the gaussian level. Physically they correspond to
Josephson plasmons in the d-wave superconductor, which were
studied in detail by Paramekanti \emph{et al}
[\onlinecite{Randeria}]. The structure of their spectrum in a
layered system is as follows. At low momenta there is a wide
regime of linear dispersion, with a small plasmon gap
${\omega}_{pc}$ due to the c-axis Josephson coupling.
Experimentally, $\omega _{pc}\approx 8-10$K in
Bi2212\cite{caxis1,caxis2,caxis3}, which is lower than typical
temperatures of interest in the normal state of lightly underdoped
cuprates. we shall therefore neglect this gap. At the high
momentum side, the linear plasmon dispersion terminates at a
characteristic energy scale $\omega _{p}$ at $q\sim\pi/\xi$. In
the cuprates ${\omega}_p$ is a high energy scale of order a few
eV. Therefore, quantum dynamics of the plasmons must be taken into
account fully at the temperatures of interest.

How do these fluctuations affect the fermion spectral function? The most
relevant coupling between the Dirac Fermions and the longitudinal phase
fluctuations is the current-current coupling\cite{Dorsey} of the form $%
\mathbf{v}_f\cdot \nabla{\varphi}\bar{\psi}_{\sigma }\psi _{\sigma }$. The
scaling dimension of this coupling is easily seen to be $-1/2$ making it
irrelevant at low energies. Further scaling arguments, and direct
calculation, show that the life-time of quasi-particles due to this coupling
diverges at low energies as $1/{\tau }\sim ({\epsilon}-{\epsilon}_F)^{2}$.
At finite temperature, this will be cut off at $T^{2}$, which gives thermal
broadening of the quasi-particles. However a Fermi arc does not form due to
coupling to the longitudinal phase fluctuations.

\emph{Transverse phase fluctuations-- } The fluctuations
considered in the calculation of the electronic Green's function
(\ref{G}) are the transverse (vortex) contribution. Vortices are
macroscopic objects whose motion is expected to be overdamped.
This picture is supported by the fact that the measured Nernst
signal and diamagnetism in underdoped cuprates\cite{Ong} are
consistent with the predictions from a classical $x-y$ model with
overdamped dynamics\cite{Podolsky}. In writing (\ref{G}) as a
static average, we assumed this dynamics to be sufficiently slow
on quasi-particle timescales.

The arc length at $T>T_c$ is of order $q_0(T)={\langle
\nabla{\varphi}_\perp^2\rangle}^{1/2}$. That is, the width of the of
the transverse current distribution, which we shall calculate within
the classical two dimensional $x-y$ model. We note that in the
equivalent coulomb
gas model, $q_0$ is related to the vortex density correlation: $q_0^2=(4\pi^2/%
\Omega)\sum_\bq {\langle n(\mathbf{q})n(-\mathbf{q})\rangle}/\mathbf{q}%
^2$.

This quantity was previously estimated\cite{Dorsey,FranzMillis} in the high
temperature limit using the Debye-H\"{u}ckel approximation\cite%
{HalperinNelson}. Here we present a different calculation of
$q_{0}\left( T\right) $, using a variational approach. Our result
coincides with the Debye-H\"{u}ckel approximation at high
temperature, but can also be used at lower temperatures, closer to
$T_{BKT}$. First we carry out the usual mapping of the coulomb gas
onto the Sine-Gordon model:
\begin{equation}
S_{SG}=\int d^{2}x\left[ \frac{1}{2}\frac{1}{4\pi ^{2}K}\left( \nabla \theta
\right) ^{2}-\frac{2y}{\xi _{c}^{2}}\cos \left[ \theta \left( \mathbf{x}%
\right) +\mu _{v}\left( \mathbf{x}\right) \right] \right]   \label{SGaction}
\end{equation}%
Here $K=J/T$ is the bare phase stiffness of the original
$x-y$ model, $y=\exp (-E_{c}/T)$ where $E_{c}$ is the vortex core
energy, and $\xi _{c}$ the core size. $\mu _{v}\left(
\mathbf{x}\right) $ is a local chemical potential that couples to
the vortices, and allows the calculation of vortex density
correlations. We shall now apply the self consistent harmonic
approximation (SCHA), which is known to give good results for the
correlations in the normal (i.e. gapped) phase \cite{Giamarchi}.
In this approach, the cosine term in (\ref{SGaction}) is replaced
by a quadratic mass term $\frac{1}{2}m^{2}{\theta }^{2}$. The mass
is determined by
minimizing the variational free energy $F_{0}+T{\langle S-S_{0}\rangle }_{0}$%
, where $F_{0}$ is the free energy of the quadratic variational action $S_{0}
$ and ${\langle \rangle }_{0}$ denotes a thermal average with respect to $%
S_{0}$. Given the solution $m\left( T\right) =\frac{\pi }{\xi _{c}}\left(
8yK\right) ^{\frac{1}{2\left( 1-\frac{\pi K}{2}\right) }}$, it is now
straight forward to compute $q_{0}$ within the variational action.
\begin{eqnarray}
q_{0}^{2}\left( T\right)  &=&\frac{(4\pi y)^{2}}{\xi
_{c}^{2}}\left( m\xi
_{c}\right) ^{2\pi K}  \nonumber \\
&&\times \int_{1}^{\infty }drr\ln r\sinh \left[ 2\pi Kf\left( m\xi
_{c}r\right) \right]   \nonumber \\
f\left( m\left\vert \mathbf{x}\right\vert \right)  &=&\int_{\left\vert
\mathbf{q}\right\vert <\frac{\pi }{\xi _{c}}}\frac{d^{2}q}{2\pi }\frac{e^{i%
\mathbf{q}\cdot \mathbf{x}}}{\mathbf{q}^{2}+m^{2}},  \label{q0_SCHA}
\end{eqnarray}%
Eq. (\ref{q0_SCHA}) can be shown to be of the Debye-H\"{u}ckel form
in the high temperature limit.

The parameters $J$, $E_{c}$, and $\xi_c$ that control the
temperature dependence of $q_0$, are not easily connected to
observable properties of the cuprates. The parameter $J$ is the
{\em bare} superfluid stiffness of the $x-y$ model. Nevertheless,
in a d-wave superconductor it is expected to be temperature
dependant $J(T)=J(0)-AT$  due to physics that lies out side of the
pure $x-y$ model, namely depletion of the condensate by
quiasiparticles at the gap nodes\cite{LeeWen}. In the normal
phase, the temperature dependence may well be more complicated
because of the emergent finite density of states at zero energy.
However, to leading order in the current fluctuations we may
assume that the linear decrease of $J$ persists in the normal
state. We take the parameter $A$ from experiments, that measure
the leading temperature dependence of the superfluid
density\cite{Boyce}. Note the distinction between $J(T)$ which is
the bare stiffness of the $x-y$ model and the macroscopic
stiffness, which vanishes in the normal state due to proliferation
of vortices.

\begin{figure}[t]
\centering \includegraphics[scale=0.4]{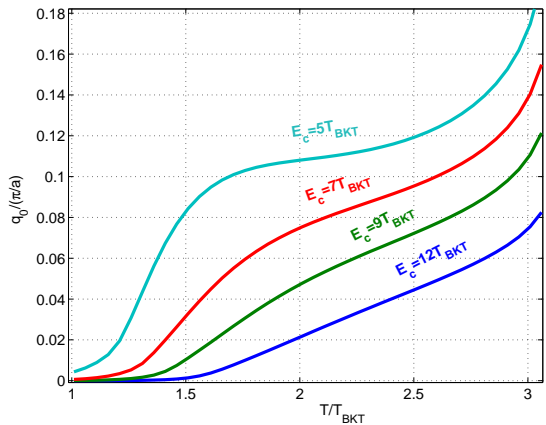} \caption{The Fermi arc length
is related to the width $q_0$ of the current distribution. This is plotted
as a function of temperature using eq. (\ref{q0_SCHA}), for several
values of $E_c/T_{BKT}$ (see text for details).} \label{fig:q0_T}
\end{figure}

The results of the calculation of the arc length (\ref{q0_SCHA})
are plotted in Fig. \ref{fig:q0_T} for a range of values of
$E_c/T_{BKT}$. We assumed that $\xi_c=5a$. Near $T_{BKT}$,
$q_{0}\left( T\right)$ is exponentially suppressed. At higher
temperatures, there is a region where $q_0$ increases rapidly due
to the proliferation of free vortices, followed by a roughly
linear increase region. Since we have assumed that $J(T)=J(0)-AT$,
the bare stiffness vanishes at some temperature (which is about
$3T_{BKT}$ for our chosen value of $A=0.3$). At this temperature
$q_0(T)$ diverges logarithmically, and the continuum description
of the $x-y$ model breaks down. The temperature dependence of the
bare stiffness $J$ above $T_c$ in the cuprates is not clear, even
though there is some evidence that it continues to decay linearly
over a considerable range\cite{Corson}.

In our treatment, we have neglected the inter-plane Josephson
coupling. This coupling becomes relevant at some $T_{c}>T_{BKT}$,
where a phase transition to a three dimensional ordered state
occurs. At this temperature, our treatment of a two dimensional
$x-y$ model is no longer valid, vortex formation is strongly
suppressed, and we expect $q_{0}\left( T\right) $ to drop abruptly
to zero. Indeed, the observed arc length seems to drop to zero at
$T_c$ [\onlinecite{Kanigel06}].

{\em Discussion and conclusions --} Before concluding let us
remark that, we have so far not payed special attention to effects
of proximity to the Mott insulating state at half filling. The
simplest (though not necessarily the correct) way to include the
effect of the strong local repulsion, is via the slave boson mean
field theory. Within this approach, the current carried by a
quasiparticle is renormalized by a factor $\a$, which is
proportional to the hole doping $x$ [\onlinecite{LeeWen}]. Hence
the doppler shift of the quasiparticle dispersion (\ref{Eq}) is
renormalized by a similar factor, so that $\bv_f\cdot \bq \to
\a\bv_f\cdot \bq$. This would have a dramatic effect on our
earlier considerations. In particular, application of a static
current with $\bq$ parallel to $\bv_f$ would no longer leave
invariant the point of zero energy excitations. Accordingly, the
average over the current distribution (\ref{Aq}), would not result
in a singular peak on the Fermi surface. Specifically, we find the
leading edge of the EDC on the Fermi surface behaves as $|\w|$ in
this case. It is interesting to note that the same renormalization
factor $\a$, which leads to the present discrepancy of slave boson
theory with experiment, is also responsible for its famous
failure\cite{LeeWen} to explain the doping independent slope of
the superfluid density with temperature\cite{Boyce}.

To summarize, we have shown that destruction of a $d$-wave
superconductor by proliferation of {\em thermal} vortices must be
accompanied by formation of at least a partial fermi surface. We
argued that this phenomenon is fundamental to the appearance of
Fermi arcs in the normal state of underdoped
cuprates\cite{Norman98}, and may partially explain their evolution
with temperature\cite{Kanigel06}. The main results rely on two
central assumptions.  The first is that vortices are purely
thermal and their quantum dynamics may be neglected. The second is
the calculation the fermion spectral function in the presence of
static vortices within a semiclassical approximation along the
lines of [\onlinecite{Volovik}]. The semiclassical approximation
was tested in numerical simulations\cite{Marinelli}, which
verified that it is controlled by the small parameter $v_\D/v_f$.
It therefore seems safe to apply the semiclassical approach to the
high $T_c$ cuprate superconductors, except in the very underdoped
regime.

In extremely underdoped materials quantum dynamics of the vortices is also expected
to be increasingly important, in violation of our first assumption. Indeed, the onset of
superconductivity possibly corresponds to a quantum phase transition, driven by
fluctuating quantum vortices\cite{Tesanovich,Sachdev}. An important open question is how the normal state we describe, that
arises in a thermal vortex liquid, evolves to the highly quantum regime near the critical doping
for the onset of superconductivity.

{\bf Acknowledgements} We are grateful to E. Demler, A. Kanigel, S.
Kivelson, A. Paramekanti, and D. Podolsky for illuminating discussions. This
research was supported in part by the NSF under Grant No.
PHY05-51164 and the U.S.-Israel binational science foundation.

\begin{widetext}

\begin{center}
\Large{\bf Supplementary information}
\end{center}

\section{Derivation of the Angle resolved Photoemission (ARPES) weight}

\subsection{Static current}
A static current is parameterized by a wave-vector $\bq$ that can
be ascribed to a twist in the order parameter, or to the hopping
matrix elements of the Hamiltonian as an external vector
potential. The two descriptions are related by a simple gauge
transformation. The ARPES spectrum in absence of perpendicular
magnetic field is given by the gauge invariant Green's function
\be I(\bk,\w)= i\int d\br dt
G^<_{\s\s}(\br,t)\exp\left(+i\bk\cdot\br +i\w t -i\int_{0}^{\br}
{\bf A}\cdot{\bf dl}\right) \label{Gless} \ee where
$G^<_{\s\s}(\br,t)=\av{c\yd_{\s,0}(0)c\nd_{\s,\br}(t)}$. We induce
a current using the gauge ${\bf A}=\bq$, for which \be
I_\bq(\bk,\w)=\int dt e^{i\w t} \av{c\yd_{\s,\bk-\bq}(0)
c\nd_{\s,\bk-\bq}(t)} \label{Gless} \ee Note that this is simply
the equilibrium expression for $G^<$, boosted by momentum $\bq$.
We express (\ref{Gless}) in terms of the Bogoliubov operators
$\a_\bk$ that diagonalize the Hamiltonian: \bea
I_\bq(\bk,\w)&=&|u_{\bk-\bq}|^2\av{\a\yd_{\bk-\bq,\ua}
\a\nd_{\bk-\bq,\ua}}\d(\w-E_{\bk-\bq}(\bq))
+|v_{\bk-\bq}|^2\av{\a\nd_{-\bk+\bq,\da}
\a\yd_{-\bk+\bq,\da}}\d(\w+E_{\bq-\bk}(\bq))\nn\\
&=&n_F(\w)\Big(|u_{\bk-\bq}|^2\d(\w-E_{\bk-\bq}(\bq))
+|v_{\bk-\bq}|^2\d(\w+E_{\bq-\bk}(\bq))\Big)\nn\\
&=&n_F(\w)A_{\bq}(\bk,-\w). \label{Ikw} \eea Here $
E_\bk(\bq)={\bf v}_f(\bk)\cdot\bq+\sqrt{\xi_\bk^2+\D_\bk^2} $ is
the quasi-particle spectrum to leading order in the current. Note
that the coherence factors $u_\bk$ and $v_\bk$ are unchanged by
the current to this order in $\bq$. That is,
$|u_\bk|^2=(1+\xi_\bk/E_\bk(0))/2$ and
$|v_\bk|^2=(1-\xi_\bk/E_\bk(0))/2$.

\subsection{Fluctuating current}
As we discuss in the paper, the spectral function in the presence
of the phase fluctuations is obtained by averaging (\ref{Ikw})
over a distribution of static currents. This leads to the
following integral: \bea A(\bk,\w)={1\over 2}\int d\h d\z
P(\h,\z)\sum_{s=\pm 1}\left(1+{s(\xi-\h)\over\sqrt{(\xi-\h)^2
+(\D-\z)^2}}\right)\d(\w-\h- s\sqrt{(\xi-\h)^2+(\D-\z)^2}),
\label{Aintegral} \eea where we have denoted:
$\D\equiv\mathbf{v}_\D\cdot (\mathbf{k}-\mathbf{k}_n)$, $\xi\equiv
\mathbf{v}_f\cdot
(\mathbf{k}-\mathbf{k}_n)$, ${\eta}=\mathbf{v}_f\cdot\mathbf{q}$, and $\zeta=\mathbf{%
v}_\D\cdot\mathbf{q}$. The current distribution is assumed to be
gaussian, such that \be P(\h,\z)={1\over\pi v_f v_\D
q_0^2}\exp\left(-{\h^2\over v_f^2 q_0^2} -{\z^2\over v_\D^2
q_0^2}\right) \ee To compute (\ref{Aintegral}) we change to the
polar coordinates $r\cos\t=\xi-\h$, $r\sin\t=\D-\z$. Then
(\ref{Aintegral}) takes the form \be A(\bk,\w)=\half\int d\t dr r
{\tilde P}_{\xi,\D}(r,\t)\sum_{s=\pm1} (1-s\cos\t)\d\left(\w-\xi+s
r(1+s \cos\t)\right) \label{Aint2} \ee where ${\tilde
P}_{\xi,\D}(r,\t)\equiv P(\xi-r\cos\t,\D-r\sin\t)$. Integrating
over $r$ to resolve the delta functions and changing variables to
$\a=s\cos\t$ we have \bea A(\bk,\w)=\half\int_{-1}^1 d\a
{(1-\a)^{1/2}\over(1+\a)^{5/2}}
\left\{(\w-\xi)\Theta(\w-\xi)F_{\xi,-\D}(\w-\xi,\a)
+(\xi-\w)\Theta(\xi-\w) F_{\xi,\D}(\w-\xi,\a)\right\},
\label{Aint3} \eea where \be F_{\xi,\D}(\w-\xi,\a)={1\over \pi v_f
v_\D q_0^2}\exp\left\{-{\left((\w-\xi){\a\over 1+\a}+\xi\right)^2
\over v_f^2 q_0^2}-{\left((\w-\xi)\sqrt{1-\a \over
1+\a}+\D\right)^2\over v_\D^2 q_0^2}\right\}. \label{Fkw} \ee For
wave-vectors $\bk$ on the Fermi surface we plug $\xi=0$ into
(\ref{Aint3}) and (\ref{Fkw}). In this case
$F_{0,-\D}(\w,\a)=F_{0,\D}(-\w,\a)$ and therefore (\ref{Aint3}) is
reduced to \be A(\bk,\w)={|\w|\over 2}\int_{-1}^1
d\a{(1-\a)^{1/2}\over(1+\a)^{5/2}}F_{0,\D}(|\w|,\a) \label{AintD}
\ee The divergence at $\a=-1$ is cutoff by the first term in the
exponent of the distribution $F_{0,\D}(|\w|,\a)$ which leads to
the singular contribution at zero frequency: \be
A_{sing}(\bk,\w)=(v_\D q_0)^{-1}e^{-\D_\bk^2/v_\D^2
q_0^2}\sqrt{v_f q_0\over |\w|} \label{Asing} \ee on the other
hand, on the nodal line ($\D_\bk=0$) (\ref{Aint3}) simplifies to
\be A(\bk,\w)=\half |\w-\xi|\int_{-1}^1 d\a{(1-\a)^{1/2} \over
(1+\a)^{5/2}}F_{\xi,0}(\w-\xi,\a) \label{Aintxi} \ee Again the
divergence of the integral is cut off by the distribution
function. Now the peak is at non vanishing frequency and it
disperses a $\w=\xi_\bk$ \be A_{sing}(\bk,\w)\approx{1\over v_\D
q_0}\sqrt{v_f q_0\over |\xi-\w|} \ee

\section{Role of longitudinal quantum phase fluctuations}

In order to estimate the effect of longitudinal quantum phase
fluctuations on the low energy fermion properties, we consider the
following model of two dimensional nodal Dirac fermions coupled to
gaussian phase fluctuations:
\begin{equation}
S=S_{F}+S_{\theta }+\lambda S_{I}
\end{equation}
\begin{equation}
S_{F}=\sum_{j}\int d\tau \int d^{2}x\left[ \bar{\Psi}_{j}\left(
\partial _{\tau }-\sigma ^{z}\mathbf{v}_{F}\cdot
\frac{1}{i}\mathbf{\nabla }-\sigma
^{x}\mathbf{v}_{\Delta }\cdot \frac{1}{i}\mathbf{\nabla }\right) \Psi _{j}%
\right]
\end{equation}
\begin{equation}
S_{\theta }=\int d\tau \int d^{2}x\left[ \frac{1}{V_{0}}\left(
\partial
_{\tau }\theta \right) ^{2}+\Lambda _{0}\left( \nabla \theta \right) ^{2}%
\right]
\end{equation}
\begin{equation}
S_{I}=\int d\tau \int d^{2}x\sum_{j}\left[
-\frac{\mathbf{v}_{F,j}}{2}\cdot \mathbf{\nabla }\theta
\bar{\Psi}_{j}\Psi _{j}\right]   \label{S_I}
\end{equation}
Here $\Psi _{j}=\left( \psi _{j,+,\uparrow
},\bar{\psi}_{j,-,\downarrow }\right) $ is the Nambu spinor
related to the $j$th pair of nodes at $\pm \bk_j$ (hence the index
$\pm$ of $\psi$).
 $\lambda$ is a small parameter. The phase action
$S_{\theta }$ is taken to be strictly in 2d, rather than a layered
system. This simplification does not change the results below,
since the essential property of $S_{\theta }$ is the linear
2d-like dispersion near the origin. As we discussed in the paper,
the full layered dispersion is nearly linear near the origin
(except for a small gap at $\mathbf{k}=0$ due to the c-axis
Josephson coupling, which is negligible at the temperatures of
interest here).

$S_I$ is the minimal coupling action of the fermions to the phase
field, of which we keep only the most relevant term consisting of
a current-current coupling.

We will treat $\lambda S_{I}$ perturbatively. The "engineering"
scaling dimensions of the fields can be read off from the fixed
point action with $\lambda =0$: $\left[ \Psi _{j}\right] =-1$,
$\left[ \theta \right] =-\frac{1}{2}$. Therefore the
scaling dimension of $S_{I}$ is found to be $\left[ S_{I}\right] =-\frac{%
1}{2}+O\left( \lambda \right) $, i.e. it is irrelevant in the weak
coupling limit. Physically, this means that the coupling to
longitudinal phase fluctuations does not change the low energy
spectrum of the fermions. In particular, we can estimate the
quasiparticle lifetime due to $S_{I}$. To leading order, $1/\tau
\left( E\right) \sim \lambda ^{2}$. In order to determine the
energy dependence, we can perform an RG transformation that takes
$E$ to some fixed energy scale $E_{0}>E$. Under this
transformation, \be
\lambda \rightarrow \lambda ^{\prime }=\left( \frac{E_{0}}{E}\right) ^{-%
\frac{1}{2}+O\left( \lambda \right) }\lambda \ee therefore
\be\frac{1}{\tau \left( E_{0}\right) }\sim (\lambda^\prime)^2
=\left( \frac{E_{0}}{E}\right) ^{-1+O\left( \lambda \right)
}\lambda ^{2}.\ee Now, since $1/\tau$ has units of energy, we can
scale back to get \be \frac{1}{\tau \left( E\right)
}=\frac{E}{E_{0}}\frac{1}{\tau \left( E_{0}\right) }\sim \left( \frac{E}{%
E_0}\right) ^{2-O\left( \lambda \right) }\lambda ^{2}\ee We
conclude that for weak coupling $1/ \tau \left( E\right) \sim
E^{2}$, and the low energy quasiparticles are well defined. A
direct evaluation of the leading diagram for $1/{\tau \left(
E\right) }$ confirms this result.

\section{Calculation of the typical current fluctuation}

The action for the transverse (vortex) part of the phase field in the 2d $%
x-y $ model is the form of a Coulomb gas model, which can be
mapped onto the Sine-Gordon model

\begin{equation}
S_{SG}=\int d^{2}x\left[ \frac{1}{2}\frac{1}{4\pi ^{2}K}\left(
\nabla \theta
\right) ^{2}-\frac{2y}{\xi _{c}^{2}}\cos \left[ \theta \left( \mathbf{x}%
\right) +\mu _{v}\left( \mathbf{x}\right) \right] \right]
\label{Ssg}
\end{equation}

Here $\mu _{v}$ is a vortex chemical potential term, that enables
a calculation of vortex density correlations. In particular, the
typical current fluctuation is given by
\begin{equation}
q_{0}^{2}=\left\langle \mathbf{\nabla }\varphi _{\perp }^{2}\left(
0\right) \right\rangle =\frac{4\pi ^{2}}{\Omega
}\sum_{\mathbf{q}}\frac{\left\langle
n\left( \mathbf{q}\right) n\left( -\mathbf{q}\right) \right\rangle }{\mathbf{%
q}^{2}}  \label{q_0_sq}
\end{equation}%
where $n\left( \mathbf{q}\right) $ is the Fourier transformed
vortex density. The vortex density correlation function in real
space can be expressed as:
\begin{eqnarray}
\left\langle n\left( \mathbf{q}\right) n\left( -\mathbf{q}\right)
\right\rangle &=&\frac{1}{\xi _{c}^{2}}\int \frac{d^{2}x}{\xi _{c}^{2}}e^{-%
\mathbf{q}\cdot \mathbf{x}}\left\langle n\left( \mathbf{x}\right)
n\left(
0\right) \right\rangle  \nonumber \\
\left\langle n\left( \mathbf{x}\right) n\left( 0\right)
\right\rangle
&=&-\left( 2y\right) ^{2}\left\langle \sin \left[ \theta \left( \mathbf{x}%
\right) \right] \sin \left[ \theta \left( 0\right) \right]
\right\rangle +2y\xi _{c}^{2}\delta \left( \mathbf{x}\right)
\left\langle \cos \left[ \theta \left( 0\right) \right]
\right\rangle  \label{n_n_correlation}
\end{eqnarray}
The last line is obtained by taking $\mu _{v}$ derivatives of the
Sine-Gordon partition function. Charge neutrality of the Coulomb
gas model fixes $\int d^{2}x\left\langle n\left( \mathbf{x}\right)
n\left( 0\right) \right\rangle =0$. Indeed, we see that otherwise
$q_{0}^{2}$ (which is proportional to the total electostatic
energy) diverges.

In order to calculate (\ref{n_n_correlation}) in the gapped
(disordered)
phase, we use the self consistent harmonic approximation. The action (\ref%
{Ssg}) is replaced by the quadratic action
\begin{equation}
S_{SCHA}=\frac{1}{4\pi ^{2}K}\int d^{2}x\left[ \frac{1}{2}\left(
\nabla \theta \right) ^{2}+\frac{m^{2}}{2}\theta ^{2}\right]
\label{S_SCHA}
\end{equation}
$m$ is a variational parameter determined by minimizing the free
energy of
the system. The optimal value is $m=\frac{\pi }{\xi _{c}}\left( 8yK\right) ^{%
\frac{1}{2\left( 1-\frac{\pi K}{2}\right) }}$. The calculation of (\ref%
{n_n_correlation}) with the action (\ref{S_SCHA}) is strait
forward, and yields
\begin{eqnarray}
\left\langle n\left( \mathbf{x}\right) n\left( 0\right)
\right\rangle _{0} &=&-\left( 2y\right) ^{2}\left( \xi
_{c}m\right) ^{2\pi K}\sinh \left[ 2\pi Kf\left( m\left\vert
\mathbf{x}\right\vert \right) \right] +A\xi _{c}^{2}\delta \left(
\mathbf{x}\right)\nn\\
f\left( m\left\vert \mathbf{x}\right\vert \right)
&=&\int_{\left\vert \mathbf{q}\right\vert <\frac{\pi }{\xi _{c}}}\frac{d^{2}q}{%
2\pi }\frac{e^{i\mathbf{q}\cdot \mathbf{x} }%
}{\mathbf{q}^{2}+m^{2}} \label{n1n2_SCHA}
\end{eqnarray}
Here $A=2y\left\langle \cos \left[ \theta \left(
\mathbf{x}_{1}\right) \right] \right\rangle $. Instead of
calculating $A$ directly, we will adjust it so that
(\ref{n1n2_SCHA}) satisfies the exact
charge neutrality sum rule $\int d^{2}x\left\langle n\left( \mathbf{x}%
\right) n\left( 0\right) \right\rangle =0$. In the high temperature ($%
K\rightarrow 0$) limit, (\ref{n1n2_SCHA}) reduces to the
Debye-H\"{u}ckel form.

Plugging (\ref{n1n2_SCHA}) into (\ref{q_0_sq}) and using $\int \frac{d^{2}q}{%
2\pi }\frac{e^{i\mathbf{q\cdot x}}}{\mathbf{q}^{2}}=\ln
\frac{L}{\left\vert \mathbf{x}\right\vert }$ (where $L$ is the
system size, which is taken to infinity at the end of the
calculation), we finally get
\begin{eqnarray}
q_{0}^{2}\left( T\right) =\frac{\left( 4\pi y\right) ^{2}}{\xi _{c}^{2}}%
\left( \xi_c m\right) ^{2\pi K}
 \int_{1}^{\infty}drr\ln r\sinh \left[ 2\pi Kf\left( m\xi _{c}r\right) %
\right] \label{q0_SCHA}
\end{eqnarray}

\end{widetext}
\end{document}